 \documentclass[final,5p,times,twocolumn]{elsarticle}

\usepackage{amsmath}
\usepackage{amssymb}

\newcommand{\nn}{\nonumber} 
\newcommand{\bn}{{\bar n}}

\newcommand{\mcdot}{\!\cdot\!}

\newcommand{\be}{\begin{equation}}
\newcommand{\ee}{\end{equation}}

\newcommand{\SCETb}{\mbox{${\rm SCET}_{\rm II}$ }}

\newcommand{\vect}[1]{\mathbf{#1}}
\newcommand{\abs}[1]{\left\lvert #1\right\rvert}
\newcommand{\bra}[1]{\left\langle #1\right\rvert}
\newcommand{\ket}[1]{\left\lvert #1\right\rangle}

\newcommand{\minus}{\!-\!}
\newcommand{\plus}{\!+\!}

\newcommand{\nkll}{N$^k$LL\ }
\newcommand{\as}{\alpha_s}
\newcommand{\MSbar}{\overline{\text{MS}}}

\newcommand{\eg}{\emph{e.g.},\ }

\newcommand{\cO}{\mathcal{O}}
\newcommand{\cM}{\mathcal{M}}

\newcommand{\cI}{\mathcal{I}}
\newcommand{\cT}{\mathcal{T}}

\newcommand{\cA}{\mathcal{A}}
\newcommand{\cS}{\mathcal{S}}

\newcommand{\cP}{\mathcal{P}}
\newcommand{\cY}{\mathcal{Y}}

\newcommand{\cV}{\mathcal{V}}
\newcommand{\cR}{\mathcal{R}}

\newcommand{\e}{\epsilon}
\newcommand{\ve}{\varepsilon}

\newcommand{\eq}[1]{Eq.~\eqref{eq:#1}}
\newcommand{\eqs}[2]{Eqs.~\eqref{eq:#1} and \eqref{eq:#2}}
\newcommand{\eqss}[3]{Eqs.~\eqref{eq:#1}, \eqref{eq:#2}, and \eqref{eq:#3}}
\renewcommand{\sec}[1]{Sec.~\ref{sec:#1}}
\newcommand{\ssec}[1]{Sec.~\ref{ssec:#1}}

\newcommand{\appx}[1]{\ref{appx:#1}}
\newcommand{\fig}[1]{Fig.~\ref{fig:#1}}
\newcommand{\figs}[2]{Figs.~\ref{fig:#1} and \ref{fig:#2}}
\newcommand{\figss}[3]{Figs.~\ref{fig:#1}, \ref{fig:#2}, and  \ref{fig:#3}}

\newcommand{\eec}{$ee$}
\newcommand{\ep}{$ep$}
\newcommand{\pp}{$pp$}
\newcommand{\eeeppp}{\mbox{$ee,\,ep,\,{\rm and}\, pp$ }}

\DeclareMathOperator{\Tr}{Tr}
\DeclareMathOperator{\Li}{Li}
\DeclareMathOperator{\PV}{P.\!V.}

\begin{document}

\begin{frontmatter}



\title{Equality of hemisphere soft functions for $e^+e^-$, DIS and $pp$ collisions at $\cO(\as^2)$}

\author[a]{Daekyoung Kang}
\ead{kang1@lanl.gov}
\author[b,c]{Ou Z. Labun}
\ead{ouzhang@email.arizona.edu}
\author[a]{Christopher Lee}
\ead{clee@lanl.gov}
\address[a]{Theoretical Division, MS B283, Los Alamos National Laboratory, Los Alamos, NM  87545, USA}
\address[b]{Department of Physics, University of Arizona, Tucson, AZ 85721, USA}
\address[c]{Institut de Physique Nucl\'eaire, IN2P3-CNRS, Universit\'e Paris-Sud, 91406 Orsay Cedex, France}


\begin{abstract}
We present a simple observation about soft amplitudes and soft functions appearing in factorizable cross sections in $ee$, $ep$, and $pp$ collisions that has not clearly been made in previous literature, namely, that the hemisphere soft functions that appear in event shape distributions in $e^+e^-\to\text{dijets}$, deep inelastic scattering (DIS), and in Drell-Yan (DY) processes are equal in perturbation theory up to $\cO(\as^2)$, even though individual amplitudes may have opposite sign imaginary parts due to changing complex pole prescriptions in eikonal propagators for incoming vs. outgoing lines. We also explore potential generalizations of this observation to soft functions for other observables or with more jets in the final state. 
\end{abstract}


\end{frontmatter}



\section{Introduction}
\label{sec:intro}

The high precision computation in Quantum Chromodynamics (QCD) of cross sections containing jets relies heavily on factorization to organize the necessary perturbative computations and accounting of nonperturbative effects \cite{Collins:1989gx,Sterman:1995fz}. Jet production in $e^+e^-$ collisions, in deep inelastic scattering (DIS), or $pp$ collisions involves physics at hierarchically separated scales of the hard collision/production of partons, of collinear splittings and emissions, of soft radiation between energetic collinear partons, and of confinement/hadronization. Factorization of the physics at these scales allows for resummation of large logarithms of scale ratios in perturbative expansions  \cite{Contopanagos:1996nh} and of rigorous proof of universality of nonperturbative effects \cite{Lee:2006fn,Lee:2006nr,Mateu:2012nk}.

In this paper we focus on soft functions describing the soft radiation between collinear jets/beams in $e^+e^-$ collisions, DIS, and Drell-Yan (DY) processes. Factorization theorems in these processes take the generic form, for $n_B$ incoming hadronic beams and $N$ outgoing jets,
\be
\label{eq:factorization}
\sigma = \Tr(H S) \otimes  J_1\otimes \cdots J_N \otimes \prod_{i=1}^{n_B} B_i \,,
\ee
where $H,S$ are hard and soft functions which are, in general, matrices in the space of color channels available in the process. The $\otimes$ signify convolutions of the beam, jet, and soft functions, whose exact form depends on the observable being measured in $\sigma$. We will distinguish soft functions for \eeeppp collisions as $S^{ee}$, $S^{ep}$, and $S^{pp}$. 

The main class of observables we are motivated to study is event shapes $\tau$ that isolate events with collinear particles in two separate (outgoing or incoming) directions when $\tau\ll 1$, \eg thrust in $e^+e^-$ \cite{Farhi:1977sg}, 1-jettiness \cite{Kang:2012zr,Kang:2013nha} or DIS thrust \cite{Antonelli:1999kx} in $ep$ collisions, and 0-jettiness or beam thrust in DY \cite{Stewart:2009yx,Stewart:2010pd,Stewart:2010tn}. For two collinear directions, $H,S$ in \eq{factorization} are numerical valued functions, the color space being 1-dimensional.

The hemisphere soft function $S_2^{ee}(\ell_1,\ell_2)$ for dijets in $e^+e^-$ has been computed in perturbation theory up to $\cO(\as^2)$ \cite{Kelley:2011ng,Monni:2011gb,Hornig:2011iu}. It is a function of $\ell_{1,2}= n\cdot k_s^{R}$  ($\bn\cdot k_s^L$), the smaller light-cone component of momentum $k_s$ of soft particles in the right (left) hemisphere with respect to the thrust axis $\vect{\hat z}$ of two back-to-back jets in the directions  $n = (1,\vect{\hat{z}})$ and $\bn = (1,-\vect{\hat z})$. Together with the $\cO(\as^2)$ hard function \cite{Moch:2005id,Idilbi:2006dg, Becher:2006mr} and collinear jet function  \cite{Becher:2006qw}  (and  $\cO(\as^3)$  anomalous dimension \cite{Becher:2006mr,Moch:2004pa}), the $\cO(\as^2)$  soft function provides enough information to predict $e^+e^-$ dijet event shapes to an unprecedented N$^3$LL  accuracy (see, \eg \cite{Abbate:2010xh,Almeida:2014uva} for definition of \nkll accuracy), which together with fixed-order N$^3$LO results, has led to the most precise extractions to date of the strong coupling $\as$ and leading nonperturbative moment $\Omega_1$ from data on event shapes \cite{Abbate:2010xh,Hoang:2015hka}.

Event shape cross sections in DIS and DY, however, have not yet reached this level of accuracy, in part due to the absence of a similar computation of the relevant soft functions to $\cO(\as^2)$. The hard and jet functions that appear in the factorization theorems are the same, but the soft functions could, in principle, be different. DIS and DY factorization theorems also contain beam functions, which have only recently been computed to $\cO(\as^2)$ \cite{Gaunt:2014cfa,Gaunt:2014xga,Gaunt:2014xxa}. This makes the $\cO(\as^2)$ soft functions $S_2^{ep,pp}$ the last remaining ingredient needed for N$^3$LL accuracy in resummation of DIS and DY event shapes. (The $\cO(\as^2)$ soft function for $k_T$-dependent distributions in DY has been computed in  \cite{Li:2011zp}.)

The difference in \eeeppp soft functions is in the direction of the path of the Wilson lines appearing in the matrix elements that define them, \eg
\be
\label{eq:YinYout}
\begin{split}
Y_n^{+\dag}(x) &= P\exp\biggl[ig \int_0^\infty ds\,n\cdot A_s(ns+x) \biggr] \\
Y_{n}^-(x) &= P\exp\Bigl[ig\int_{-\infty}^0 ds\,n\cdot A_s(n s + x)\Bigr] \,,
\end{split}
\ee
where $A_s = A_s^A T^A$, $T^A$ being the generators in the fundamental representation of SU(N). 
In $Y_n^+$, $n$ is the direction of an outgoing jet in $ee$ or $ep$, while in $Y_n^{-}$ it is the direction of an incoming hadron beam in $ep$ or $pp$. Feynman rules for gluons emitted from the two Wilson lines in \eq{YinYout} are the same except for the sign of $i\e$ in the eikonal propagators determining the complex pole prescription. For example, the amplitudes for emission of a gluon of momentum $k$  from the eikonal lines in \eq{YinYout} are
\be
\cA_{1n}^+  = -g\mu^\epsilon \frac{n\cdot \varepsilon(k)}{n\cdot k + i\epsilon}  \,, \quad \cA_{1n}^- = -g\mu^\epsilon \frac{n\cdot \varepsilon(k)}{n\cdot k - i\epsilon} \,,
\ee
where $\varepsilon(k)$ is the polarization vector for an outgoing gluon. These differences in soft Wilson lines appearing in factorization theorems for cross sections with incoming or outgoing collinear particles were studied extensively in \cite{Chay:2004zn,Arnesen:2005nk}. 
This subtle difference is enough to potentially change the result of perturbative computations. Ignorance of whether this actually occurs or not has so far been the roadblock to N$^3$LL accuracy in resumming DIS and DY event shapes. (Nonperturbatively, the three soft functions must be assumed to be different.)

In this paper, we compare all the perturbative amplitudes that could appear in the computation of the \eeeppp soft functions up to $\cO(\as^2)$. The amplitudes themselves are not dependent on the observable being measured in the final state, so our conclusion is fairly generally applicable. We find that nearly all amplitudes are transparently equal whether the particles originate from incoming or outgoing Wilson lines. The exception is a subset of the $\cO(g^3)$ 1-gluon emission amplitudes, namely, those 1-loop amplitudes containing a triple gluon vertex [($2\cT$) in \fig{1loop}], which is part of the computation of the soft gluon current at one loop \cite{Catani:2000pi} (and computed to two loops in \cite{Li:2013lsa}).
For $ee$ and $ep$ these amplitudes are equal, but for $pp$ it has the opposite sign in the imaginary part. These imaginary terms cancel, however, upon summing all products of amplitudes and their complex conjugates that contribute to the final soft functions.

Although this result follows immediately from existing results on the 1-loop soft gluon current, the consequent equality of the $ee$, $ep$, and $pp$ soft functions has not be made clearly in the literature and has not yet been used to extend resummation of $ep$ and $pp$ event shapes to N$^3$LL accuracy. (See, however, preliminary results, including observation about equality of soft functions, in \cite{SCET2014,QCD2014,HERA2014}.) It is one of the purposes of this letter to make this simple, though unnoticed, observation explicit. The results for the two-loop soft functions for $e^+e^-$ event shapes in \cite{Kelley:2011ng,Monni:2011gb,Hornig:2011iu} thus can be immediately used for $ep,pp$ event shapes as well. The equality of soft functions in these three different processes, furthermore, extends to many other observables besides event shapes.

In \sec{soft} we review the factorization theorems for event shapes in \eeeppp collisions in which the soft functions that we study appear. In \sec{equality} we consider all possible amplitudes that could contribute to the soft functions at $\cO(\as^2)$, in particular the one-loop real emission amplitude. We observe that those are equal for $ee$ and $ep$ but complex conjugated for $pp$, though their final contributions to the soft functions are equal. We also consider generalization to soft functions containing Wilson lines for gluon beams/jets and those with more than two legs. In \sec{conclusion} we conclude. In the appendices we summarize the final result for the hemisphere soft function, previously calculated for $e^+e^-$, and provide additional details of some of our computations.

\section{Factorization and soft functions for $ee,ep,pp$ collisions}
\label{sec:soft}

In this section, we review the contexts in which the three types of soft functions we consider in this paper appear, for two-jet event shapes in $e^+e^-$ collisions, for one-jet event shapes in DIS, and for 0-jet or beam thrust event shapes in $pp$ collisions.

A generic way to define event shapes in any of these types of collisions is in terms of $N$-jettiness \cite{Stewart:2010tn}:
\be
\label{eq:Njettiness}
\tau_N = \frac{2}{Q^2}\min\sum_{i}\{ q_a\cdot p_i,q_b\cdot p_i,q_1\cdot p_i,\dots,q_N\cdot p_i\}\,,
\ee
where $Q$ is the hard interaction scale and the $q_k$ are lightlike 4-vectors in the directions of any incoming beams $a,b$ and $N$ outgoing jets. The minimum operator groups all final-state particles $i$ into regions according to which vector $q_k$ is closest. An event with small $\tau_N\ll 1$ has $N$ well-collimated jets plus initial-state radiation (ISR) in the beam directions.

Dijet events in $e^+e^-$ collisions can be probed using global observables called \emph{event shapes} \cite{Dasgupta:2003iq}, such as  \emph{thrust} $\tau=1-T$ \cite{Farhi:1977sg,Brandt:1964sa}, corresponding to $\tau=\tau_2$ in \eq{Njettiness} with no $q_{a,b}$, and $q_{1,2} = (Q/2)(1,\pm\vect{\hat t})$,
where $Q$ is the center-of-mass energy of the collision and $\vect{\hat t}$ is the \emph{thrust axis}, the unit 3-vector that minimizes the value of $\tau$. 
Other event shapes can be defined by weighting final-state particles in the two hemispheres determined by $\vect{\hat t}$ differently, such as hemisphere masses \cite{Chandramohan:1980ry,Clavelli:1979md,Clavelli:1981yh},  broadening \cite{Catani:1992jc}, and angularities \cite{Berger:2003iw}. Event shapes relative to the broadening axis were defined in \cite{Larkoski:2014uqa}, and the $C$-parameter does not refer to a particular axis at all \cite{Parisi:1978eg,Donoghue:1979vi}.

Event shapes can also be considered in DIS, $e(k) + p(P) \to X(p_X) + e(k')$, such as the 1-jettiness $\tau_1$, defined by \eq{Njettiness} with one beam direction $q_a$ and one jet direction $q_1$. There are many different ways to choose these in terms of the DIS kinematic variables; several were considered in \cite{Kang:2012zr,Kang:2013nha,Kang:2014qba}. One, called $\tau_1^b$ in \cite{Kang:2013nha}, corresponds to the DIS thrust $\tau_Q$ defined in \cite{Antonelli:1999kx,Dasgupta:2003iq}, with the choices $q_a = xP$ and $q_1 = q+xP$, where $q = k-k'$, $x = Q^2/(2P\mcdot q)$, and $Q^2 = -q^2$.
In the Breit frame this choice divides the final state into two back-to-back hemispheres.

Finally in $pp$ collisions, the observables \emph{beam thrust} \cite{Stewart:2009yx,Stewart:2010pd} or 0-jettiness $\tau_0$ \cite{Stewart:2010tn} measure the collimation of hadronic final-state particles in $pp$ collisions along the beam directions themselves. They can be used, \eg to veto jets in the central region for Drell-Yan processes $pp\to \ell^+\ell^- X$, which plays an important role in reducing QCD backgrounds in searches for Higgs or new physics particles.
Beam thrust is defined with respect to lightlike vectors $n_{a,b}$ along the incident proton directions \cite{Stewart:2010tn}, 
$q_{a,b}^\mu = \frac{1}{2} x_{a,b} E_{\text{cm}} n_{a,b}$,
where $n_{a,b}^\mu = (1,\pm\hat z) \equiv n,\bn$ in the CM frame. 
The 0-jettiness defined by \eq{Njettiness} with these vectors is related to the beam thrust $\tau_B$ defined in \cite{Stewart:2009yx,Stewart:2010pd} by $\tau_B =  \tau_0\sqrt{\mbox{$1+\vect{q}_T^2/q^2$}}$, where $q^2$ and $\vect{q}_T$ are the dilepton invariant mass and transverse momentum, respectively.

Predictions of event shapes in QCD perturbation theory exhibit logarithms $\as^n \ln^k\tau$ that become large in the endpoint region $\tau\to 0$. In this region these logs must be summed systematically to all order in $\as$ for convergent, physical results \cite{Catani:1991kz,Catani:1992ua}. Modern resummation techniques are based on factorization and renormalization group evolution, either directly in the language of perturbative QCD \cite{Contopanagos:1996nh,Luisoni:2015xha} or using the techniques of effective field theory, in this case soft collinear effective theory (SCET) \cite{Bauer:2000ew,Bauer:2000yr,Bauer:2001ct,Bauer:2001yt,Bauer:2002nz}. Both paths lead to equivalent results in principle, though particular implementations to a given order of accuracy in the literature may differ (see \cite{Almeida:2014uva}). 

The factorization approaches lead to predictions for the $e^+e^-$, DIS, or DY beam thrust distributions (see, \eg \cite{Kang:2012zr,Kang:2013nha,Stewart:2009yx,Stewart:2010pd,Stewart:2010tn,Almeida:2014uva,Berger:2003iw,Korchemsky:1999kt})
each of which takes the form of \eq{factorization}. In each case there is a hard function $H$ which is a squared Wilson coefficient from matching the QCD current $\bar q \Gamma^\mu q$ onto a SCET operator (\eg \cite{Kang:2013nha,Bauer:2002ie,Manohar:2003vb,Bauer:2003di}); $J_{n,\bn}$ are jet functions (defined in, \eg \cite{Abbate:2010xh,Bauer:2008dt} and computed to $\cO(\as)$ in \cite{Lunghi:2002ju,Bauer:2003pi} and $\cO(\as^2)$ in \cite{Becher:2006qw}) dependent on the invariant mass $t_{n,\bn}$ of the collinear jet; and $B_i$ is a beam function \cite{Stewart:2009yx,Stewart:2010qs} dependent on the transverse virtuality and/or momentum of ISR. The $\otimes$ convolutions in \eq{factorization} combine the jet/beam variables with the soft momentum $k_s$ in $S$ properly to give the value of the measured observable. 

A careful demonstration of factorization must also account for Glauber modes that potentially violate it; such arguments for particular cross sections in QCD are given in, \eg \cite{Collins:1989gx,Collins:1985ue,Collins:2011zzd}; formulating these kinds of arguments in SCET is under active development, see, \eg \cite{Bauer:2010cc,Gaunt:2014ska}, but is not our focus here. We begin with the factorization formulae in typical use for event shape cross sections in QCD and SCET (citations above) and focus on properties of the soft functions they contain.

The soft functions in \eq{factorization} for these event shapes are projections of the hemisphere soft functions,
\be
\label{eq:softprojectionee}
S(k,\mu) = \int d\ell_1 d\ell_2 \delta(k - \ell_1- \ell_2) S_2 (\ell_1,\ell_2,\mu)\,,
\ee
where the soft function on the right-hand side has two arguments, $\ell_1,\ell_2$, which are the small light-cone components of the soft radiation in either of the two hemispheres defined by the back-to-back collinear axes $n,\bn$. The soft functions are defined in terms of a matrix element of Wilson lines that arise from a field redefinition that decouples soft and collinear interactions at leading power in the SCET Lagrangian \cite{Bauer:2001yt}, leading to
\begin{align}
\label{eq:hemisoftee}
&S_2(\ell_1,\ell_2,\mu) =  \frac{1}{N_C}\Tr\sum_{i\in X_s}  \abs{\bra{X_s} T [Y_n^{\pm\dag}(0)  Y_\bn^\pm(0)] \ket{0}}^2 \\
&\times \delta\Bigl(\ell_1 \minus \sum_{i\in X_s} \theta(\bn\cdot k_i \minus n\cdot k_i) n\cdot k_i\Bigr) \delta\Bigl(\ell_2 \minus \sum_{i\in X_s} \theta(n\cdot k_i \minus \bn\cdot k_i) \bn\cdot k_i\Bigr)\,, \nn
\end{align}
where the trace is in color space, $N_C$ is the number of colors, and $T$ denotes time-ordering.
The path of the Wilson lines depends on whether $n,\bn$ are incoming or outgoing directions. 
$Y_n^{+\dag}$ and $Y_n^-$ were defined in \eq{YinYout}, and the other possibilities are obtained by taking their Hermitian conjugate and/or replacing $n\to\bn$.
For $e^+e^-$, both lines in \eq{hemisoftee} are $+$, for $pp$ they are both $-$, and for DIS they are $Y_n^{+\dag} Y_\bn^-$  \cite{Chay:2004zn,Arnesen:2005nk}. 

Parity and time-reversal symmetry can be used to flip the directions of the Wilson lines in \eq{hemisoftee} between incoming and outgoing \cite{Collins:2011zzd}, potentially relating the $e^+e^-$ and DY soft functions; however, the time-ordering prescription in \eq{hemisoftee} gets reversed \cite{Stewart:2009yx}, foiling a potential all-orders proof of equality.

The measurements of 1-jettiness in DIS or 0-jettiness in $pp$ may not necessarily divide particles in the final state into back-to-back hemispheres, but boost properties of the Wilson lines can be used in each case to express their factorization theorems in terms of the back-to-back hemisphere soft functions \cite{Kang:2013nha,Stewart:2009yx}.

The perturbative result for $S_2^{ee}$ is known up to $\cO(\as^2)$ \cite{Kelley:2011ng,Monni:2011gb,Hornig:2011iu}, quoted in \appx{pert}. The DIS and DY hemisphere soft functions differ only in the direction of the Wilson lines in \eq{hemisoftee}. Now we proceed to consider the relations among them.

\section{Equality of soft functions at $\cO(\as^2)$ }
\label{sec:equality}
In this section we show equality of the soft functions for the three cases $e^+e^-\to\text{dijets}$, DIS 1-jettiness, and $pp$ beam thrust at  $\cO(\as^2)$. Switching the direction of a Wilson line from incoming to outgoing flips the sign of the $i\e$ in the eikonal propagators formed by emission/absorption of gluons, \emph{e.g.} \eq{C1}. This could affect the value of the diagrams. Nevertheless, we show that the final soft functions remain equal up to $\cO(\as^2)$. 

First we set up some of the notation we will use in our proof. 
The perturbative computation of the soft functions in \eq{hemisoftee} can be performed either from cut diagrams with four Wilson lines with an appropriate measurement function along the cut \cite{Hornig:2009vb}, or by computing amplitudes for emission of $n=0,1,2,\dots$ particles up to the appropriate order in $\as$ and performing the phase space integrals implicit in the sum in \eq{hemisoftee}. We will take the latter approach here. 
The result of computing \eq{hemisoftee} up to $\cO(\as^N)$ in perturbation theory takes the generic form,
\be
\label{eq:softamplitudes}
\begin{split}
S_2(\ell_1,\ell_2) &= \frac{1}{N_C}\!\Tr\! \sum_{n=0}^N \!\!\int\!\! d\Pi_n 
\cM(\ell_1,\ell_2;\{k_n\}) \! \sum_{i,j}\! \cA_j^\dag(\{k_n\})\,\cA_i(\{k_n\}),
\end{split}
\ee
where $\cA_i(\{k_n\})$ is an amplitude to emit $n$ particles with momenta $k_1,\dots, k_n$. The sum over amplitudes $i,j$ goes over those pairs of amplitudes that produce the same final state with momenta $\{k_n\}$ and have total order $\as^N$. Implicitly for each product of amplitudes there is a sum over the spins or polarizations and colors of the final state particles. The trace in \eq{softamplitudes} is over products of color matrices left over in the product of amplitudes. The phase space integration measure is given by
\be
\label{eq:phasespace}
d\Pi_n = \prod_{i=1}^n \frac{d^D k_i}{(2\pi)^D} 2\pi\delta(k_i^2)\theta(k_i^0)\,,
\ee
and the measurement function $\cM$ in \eq{softamplitudes} is 
\be
\label{eq:measurement}
\cM(\ell_1,\ell_2;\{k_n\}) = 
\delta\Bigl(\ell_1 - \sum_{i=1}^n k^+_i \, \theta(k_i^- - k_i^+)\Bigr)\,  \delta\Bigl(\ell_2 - \sum_{i=1}^n  k_i^- \, \theta(k^+_i - k^-_i)\Bigr)\,,
\ee
where $k^+ \equiv n\mcdot k $ and $k^- \equiv \bn \mcdot k.$

The relevant amplitudes that can appear up in the computation of the $\cO(\as)$ and $\cO(\as^2)$ soft functions are shown in \fig{1loop}.

We will work in dimensional regularization (DR) in the $\MSbar$ scheme, although our conclusions about equality of the soft functions to $\cO(\as^2)$ are independent of these choices. One may be concerned about using DR as an IR regulator. In fact the jet and soft functions in the event shape distributions we consider are IR finite and thus independent of the IR regulator, as argued at one loop in, \eg \cite{Hornig:2009vb,Hornig:2009kv,Manohar:2006nz}, and at two loops in, \eg \cite{Idilbi:2007ff,Idilbi:2007yi}.

\subsection{One-loop soft function}

\begin{figure}
\begin{center}
\includegraphics[width=.75\columnwidth]{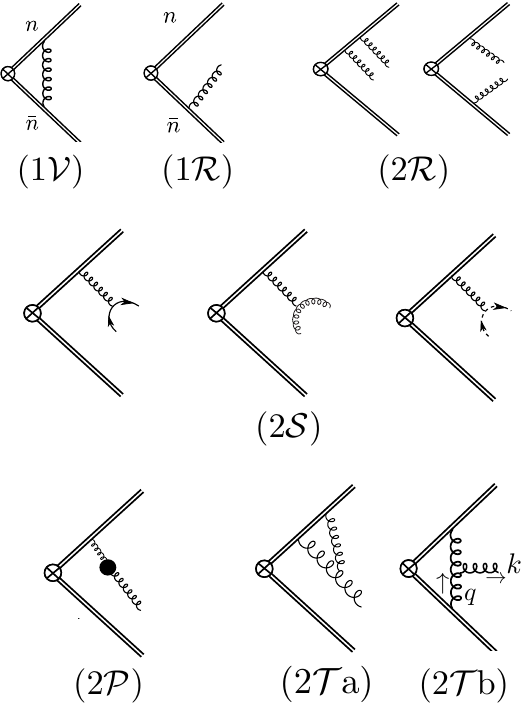}
\end{center}
\vspace{-1em}
\caption{Amplitudes contributing to the $\cO(\as)$ and $\cO(\as^2)$ soft functions: 1-gluon virtual (1$\mathcal{V}$) and real (1$\mathcal{R}$), 2 real gluon (2$\mathcal{R}$), 1-to-2 splitting (2$\mathcal{S}$), vacuum polarization (2$\mathcal{P}$), and 1-loop real gluon emission (or soft-gluon current) from a three-gluon vertex (2$\mathcal{T}$a) and (2$\mathcal{T}$b). Only (2$\cT$b) potentially differs upon changing the directions of the Wilson lines from incoming to outgoing.}
\label{fig:1loop}
\end{figure}

The one-loop result for the soft function $S_2$ can be computed from diagrams ($1\cV$) and ($1\cR$) illustrated in \fig{1loop}. There is a tree-level, 0-gluon amplitude, not drawn, which simply takes the value $\cA_0^{(0)} = 1$.
The virtual amplitude $\cA_{1\cV}$ is  scaleless and zero in dimensional regularization (DR), only playing the role of converting IR to UV divergences (\emph{e.g.} \cite{Hornig:2009kv,Manohar:2006nz}). 

The first nontrivial amplitudes are the real 1-gluon amplitudes in \fig{1loop}. For emission of a gluon of momentum $k$ from one of the two outgoing lines in \eq{hemisoftee} for $e^+e^-$,
\be
\label{eq:eikonalamplitudes}
\cA_{1R}^{n} = -g\mu^\epsilon \frac{n\cdot \varepsilon(k)}{n\cdot k + i\epsilon}  \,, \quad \cA_{1R}^{\bn} = g\mu^\epsilon \frac{\bn\cdot \varepsilon(k)}{\bn\cdot k + i\epsilon} \,,
\ee
where $\varepsilon(k)= \varepsilon^A(k) T^A$ is the polarization vector for a final-state gluon of momentum $k$.
Switching an outgoing line to an incoming line changes $+i\epsilon$ to $-i\epsilon$.
The change occurs in the amplitude $\cA_{1\cR}^{\bn} $ for $ep$ and  in both $\cA_{1\cR}^{n}$ and $\cA_{1\cR}^{\bn}$ for $pp$.
These signs are determined by the regulation of the integration limit at $\pm \infty$ in the path of the Wilson line, \emph{e.g.} ,
\be
\label{eq:C1}
\begin{split}
-ig \int_0^\infty ds\, e^{(ik\cdot \bn -\epsilon) s} &= -ig\frac{-1}{i\bn\cdot k - \epsilon} = \frac{g}{\bn\cdot k + i\epsilon}\,, \\
  ig \int_{-\infty}^0 ds\, e^{(ik\cdot \bn + \epsilon) s} &= ig\frac{1}{i\bn\cdot k + \epsilon} = \frac{g}{\bn\cdot k - i\epsilon}\,,
\end{split}
\ee
However the different $i\e$'s can be dropped because the delta function $\theta(k^0)\delta(k^2)$ ensures $k^\pm>0$ and the phase space integral does not cross the poles in the eikonal propagators. All real amplitudes for \eeeppp become the same. 
The measurement function for one real gluon is given by
\be
\cM(\ell_1,\ell_2;k) = \theta(k^- \minus k^+) \,\delta(\ell_1 - k^+) \delta(\ell_2)+\theta(k^+ \minus k^-) \,\delta(\ell_1) \delta(\ell_2 - k^-)\,,
\ee
The sum over squared amplitudes in \eq{softamplitudes} up to $\cO(\as)$ is very easily evaluated and gives the well-known result \cite{Fleming:2007xt},
\be
\label{eq:soft1loop}
S_2^{(1)}(\ell_1,\ell_2) = \frac{\as C_F}{\pi} \frac{(\mu^2 e^{\gamma_E})^{\epsilon}}{\Gamma(1-\epsilon)} \frac{1}{\epsilon} \bigl[ \ell_1^{-1-2\epsilon} \delta(\ell_2) + \ell_2^{-1-2\epsilon} \delta(\ell_1)\bigr]\,,
\ee
written in the $\MSbar$ scheme and independent of $\pm i\e$'s in \eq{eikonalamplitudes}.

\subsection{Two-loop soft function}\label{sssec:two-loop}

At $\cO(\as^2)$, an explicit computation has been given only for the $ee$ soft function \eq{hemisoftee} \cite{Kelley:2011ng,Monni:2011gb,Hornig:2011iu}. The relevant amplitudes at this order are shown in Fig.~\ref{fig:1loop}.
In DR, the 2-loop purely virtual amplitudes and 1-gluon emission amplitudes with an independent virtual loop are scaleless and zero and are not drawn.
The nonzero contributions to the $\cO(\as^2)$ soft function are given by the appropriate terms contained in \eq{softamplitudes}. The relevant contributions at this order are products of amplitudes for:
\begin{enumerate}
\item 2-real gluon emission, $\cA_{2\cR}^\dag \cA_{2\cR}$,
\item Gluon splitting to $gg$, $q\bar{q}$ and ghost pairs, $\cA_{2\cS}^\dag\cA_{2\cS}$.
\item Vacuum polarization and tree-level 1-gluon emission, $\cA_{1\cR}^{\bn\dag} \cA_{2\cP}^{n}  + (n\leftrightarrow \bn)$.
\item 1-loop single emission with a 3-gluon vertex and tree-level 1 gluon emission, $\cA_{1\cR}^{n,\bn\dag}\cA_{2\cT}$,
\end{enumerate}
and  complex conjugates. All of these have been computed in \cite{Kelley:2011ng} for $e^+e^-$, and we will not repeat the results for individual classes of diagrams but just consider their equivalence to $ep$ and $pp$. For this proof, we will actually only need to look at diagrams in category 4 in detail, and we defer this to \ssec{amplitudes}. The complete result of summing all $\cO(\as^2)$ contributions 1--4 is summarized in \appx{pert}. 

 In the derivation of the equality our proof does not 
depend on the momenta $k_i$ of the final states in \eq{softamplitudes}, nor on the measurement function 
$\cM(\ell_1,\ell_2;\{k_n\})$, but only on properties of the amplitudes $\cA_i$ themselves. Therefore, our proof
applies to various classes of observables, some of which we list in \sec{conclusion}.

It is most convenient to give results for the $\cO(\as^2)$ soft function in terms of the integrated or cumulative soft function,
\be
S_c(\ell_1,\ell_2,\mu) = \int_0^{\ell_1}\int_0^{\ell_2} d\ell_1' d\ell_2' \,S_2(\ell_1',\ell_2',\mu)\,.
\ee
The terms in the soft function at $\cO(\as^2)$ can be classified into three groups,
\be
\label{eq:S2terms}
S_c^{(2)}(\ell_1,\ell_2,\mu) = \frac{\as(\mu)^2}{4\pi^2}\Bigl[ R_c^{(2)}(\ell_1,\ell_2,\mu) + S_{\text{NG}}^{(2)}(\ell_1,\ell_2)  + c_S^{(2)} \Bigr] \,,
\ee
where $R_c$ contains $\mu$-dependent logs associated with the soft anomalous dimension, $S_{\text{NG}}$ contains the ``non-global'' terms arising from two soft gluons entering opposite hemispheres and depends non-trivially on both $\ell_1,\ell_2$ simultaneously, and the last term contains the constant $c_S^{(2)}$.

Before looking at individual diagrams, we can deduce which parts of \eq{S2terms} must be equal for the \eec, \ep, and \pp\ hemisphere soft functions. 
The logarithmic terms in $R_c^{(2)}$ in \eq{Rc2} are the same for all three soft functions, since they have the same anomalous dimension. This follows from the factorization theorem \eq{factorization} for each process in which these soft functions appear. The hard functions all have the same anomalous dimension, and the jet/beam functions all have the same anomalous dimensions. Since the cross section itself is RG-invariant ($\mu$-independent), $R_c$ must be the same for $ee,ep,pp$.

The non-global terms in $S_{\text{NG}}^{(2)}$ in \eq{SNG2} are also the same, since they are entirely determined by the graphs with two real gluons, by the arguments in \cite{Hornig:2011iu}. 
As reviewed below, at $\cO(\as^2)$ the amplitudes with two real gluons are manifestly real, and the signs of the $i\e$'s in eikonal propagators do not matter. Thus they are the same for \eeeppp\!.

The only terms that could potentially differ for the three soft functions are the constant terms in $c_S^{(2)}$ in \eq{cS2}, computed for $ee$ in \cite{Kelley:2011ng,Monni:2011gb}. By examining the complex pole structure of the Feynman diagrams that can contribute, we will find in fact that they are also the same.

\subsection{Amplitudes contributing to $\cO(\as^2)$ soft functions}\label{ssec:amplitudes}

The diagrams ($2\cS$) in \fig{1loop} all have two real gluons, quark/antiquark or ghosts in the final state.  
The eikonal propagators among \eeeppp soft functions look like  $\sim 1/(p^\pm \pm i\e)$, where $p = k_1,k_2$ or $k_1+k_2$. The onshell delta function $\delta(k_i^2) \theta(k_i^0)$ where $i=1,2$ ensures that $k_{1,2}^\pm \geq 0$, and the integrals over $k_{1,2}$ in \eq{phasespace} thus do not cross over the poles in the eikonal propagators. Thus the $i\e$'s can be dropped and these contributions are the same for \eeeppp.

The vacuum polarization diagrams ($2\cP$) in \fig{1loop} have the same eikonal propagators as the single-real-gluon graphs at $\cO(\as)$ in \fig{1loop}, and the $i\e$'s in these propagators can be dropped for the same reasons as for 2-real-gluon diagrams.
The uncut gluon propagator and any propagators in diagrams ($2\cP$) remain the same for \eeeppp soft functions. Thus these diagrams make the same contribution to all three soft functions.

Now we consider the 3-gluon vertex diagrams (2$\cT$) in \fig{1loop}.
Those diagrams involve a loop with eikonal propagators whose pole prescription changes for $ee,ep,pp$, and we will investigate this integral carefully.

In diagram ($2\cT$a) and its counterpart with $n\leftrightarrow\bn$, both virtual gluons are attached to the same eikonal line. The signs of the $i\epsilon$'s in these eikonal propagators change when flipping from incoming to outgoing lines. As observed in \cite{Kelley:2011ng}, the loop integrals associated with these diagrams are scaleless and thus zero in DR for the $ee$ soft function, when both lines are outgoing. This result is independent of the directions of the Wilson lines.

Now we turn our attention to diagram ($2\cT$b), the only case where equivalence among \eeeppp diagrams is nontrivial in DR. The amplitude is
\begin{align}
\label{eq:3gluonCamplitude}
&\cA_{2\cT b}(k) = \frac{ig^3\mu^{3\e} C_A}{2(2\pi)^D}
 \!\!\int \!\! \frac{d^D q}{q^2 \plus i\e} 
 \frac{1}{(k \minus q)^2 +i\e} \frac{1 }{ (k\minus q)^+ \pm i\e} \frac{1}{q^- \pm i\e} \nn \\
& \times 
\biggl\{ \varepsilon^-(k)  (2k- q)^+ -  \varepsilon^+(k) (k+q)^- 
     - 2{\boldmath{\mbox{$\varepsilon$}}}_\perp\mcdot (\vect{k}_\perp
     \minus 2 \vect{q}_\perp)
      \biggr\}\,,
\end{align}
where the signs of the $\pm i\e$'s in the last two propagators on the first line are $++$ for $ee$, $--$ for $pp$, and $+-$ for $ep$.
In the $\cO(\as^2)$ soft function, this amplitude will get multiplied by one of the one-gluon tree-level amplitudes in \fig{1loop}, which are proportional to $\varepsilon^+$ or $\varepsilon^-$, so in the sum over gluon polarizations in \eq{softamplitudes}, the term with ${\boldmath{\mbox{$\varepsilon$}}}_\perp$ in \eq{3gluonCamplitude} will vanish. Thus we drop it from here on.

The remaining terms in \eq{3gluonCamplitude} can be split into a scaleless, and thus zero, part and a nonzero part. The scaleless part comes from the term in numerator with $(k-q)^+$ in the first term and $q^-$ in the second, as each cancels one of the eikonal propagators on the first line. 
The nonzero part can be written
\be
\label{eq:3gluonCnonzero}
\cA_{2\cT b}(k) = \frac{i}{2}g^3\mu^{3\e} C_A  \Bigl[ \varepsilon^-(k)  k^+ -  \varepsilon^+(k) k^- \Bigr]\, \cI_\cT(k) \,,
\ee
where we have defined the integral
\be
\label{eq:ICee}
\cI_\cT \equiv  \int \!\! \frac{d^D q}{(2\pi)^D} 
\frac{1}{q^2 \plus i\e}\frac{1}{(k \minus q)^2 +i\e} \frac{1 }{(k \minus q)^+ \pm i\e} \frac{1}{q^- \pm i\e}\,.
\ee
This integral is computed explicitly in \appx{threegluon}. The result for the three cases $ee,ep,pp$ is
\be
\label{eq:3gluonresult}
\begin{split}
&\cI_{\cT} (k) = \frac{i}{16\pi^2} (4\pi)^\epsilon \Gamma(1+\e) (\vect{k}_\perp^2)^{-1-\e}  \\
&\quad \times \biggl[ \frac{2}{\e^2} - \pi^2 - 4\zeta_3\e + \frac{\pi^4}{60}\e^2 \pm i\pi \Bigl( \frac{2}{\e} - \frac{\pi^2}{3}\e - 4\zeta_3 \e^2\Bigr)\biggr] \,,
\end{split}
\ee
where the $+i\pi$ sign is for $ee,ep$ and $-i\pi$ for $pp$, consistent with the result in \cite{Catani:2000pi}. This immediately establishes for $ee$ and $ep$,
\be
\label{eq:epppequal}
\cA_{2\cT b}^{ee} = \cA_{2\cT b}^{ep}
\ee
to $\cO(g^3)$, and thus that the soft functions are equal to $\cO(\as^2)$. The differing $i\pi$ terms between $ee/ep$ and $pp$ cancel in the computation of the full soft function once we multiply by the tree-level amplitudes in \eq{eikonalamplitudes} and add complex conjugate diagrams:
\be
\begin{split}
\label{eq:differences}
\left( \cA_{2\cT b}^{ep} - \cA_{2\cT b}^{pp} \right) \cA_{1R}^{n,\bn \dag} +h.c.=0\,,
\end{split}
\ee
This establishes that the total perturbative soft functions for \eeeppp are equal up to $\cO(\as^2)$:
\be
\label{eq:finalresult}
\boxed{S_2^{(2)ee} = S_2^{(2)ep} = S_2^{(2)pp}.}
\ee
This result depends primarily on the 1-loop soft gluon current computed in \cite{Catani:2000pi} and reproduced in \eqs{3gluonCnonzero}{3gluonresult}.

\subsection{Gluon soft functions}
\label{ssec:Sgg}
Above we have discussed quark soft functions, built out of Wilson lines in the fundamental representation. 
Wilson lines for collinear gluons are defined in terms of the adjoint, \eg
\be
\cY_n^\dag(x) = P\exp\biggl[ig \int_0^\infty \!\! ds\,n\cdot A^a_s(ns+x) \,\cT^a \biggr]
\,.\label{yee}
\ee
where $(\cT^a)_{bc}=-i\, f^{abc}$.
Since the quark and gluon soft functions differ only in color factors, most of the discussion above still applies, except for possibly \eq{differences}, which relies on the color factor $i\,C_A/2$ in front of \eq{3gluonCnonzero} being purely imaginary.
The color factors for the amplitude $\cA_{2\cT b}$ for the two cases are
\be
\left.\cA_{2\cT b}\right|_\text{color} =
\ \begin{cases}
f^{ABC} T^A  T^B = \frac{i}{2}C_A \,T^C, & \mbox{for quark} \\
f^{ABC} \cT^A  \cT^B = \frac{i}{2}C_A \,\cT^C,& \mbox{for gluon}  
 \end{cases}
\ee
These factors differ only in the color matrix, which is implicit in the polarization vector $\varepsilon^\pm(k) = \varepsilon^\pm_C(k)\, T^C$ in \eq{3gluonCnonzero}. Replacing with $\varepsilon^\pm = \varepsilon^\pm_C\, \cT^C$, the amplitude for gluon Wilson lines remains in the same form. Because the color factor remains purely imaginary, the argument used to obtain \eq{differences} remains valid.
Therefore, the equality at $\cO(\as^2)$ in \eq{finalresult} is also true for the gluon soft functions.

\subsection{Multi-jet soft functions at $\cO(\as^2)$}
\label{sec:multi-jet}
\begin{figure}
\begin{center}
\center
\includegraphics[height=2.5cm]{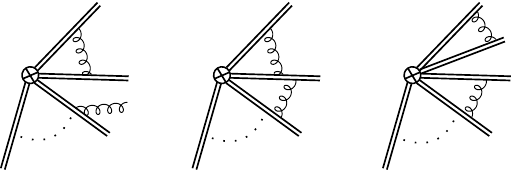}
\caption{ 
Examples of $\cO(\as^2)$ diagrams with 3 Wilson lines and 4 Wilson lines involved.
Dots represent Wilson lines not emitting soft gluons.}
\label{fig:34lines}
\end{center}
\end{figure}

Now we extend our discussion on \eeeppp soft functions to multi-jet soft functions defined in terms of more than two distinct collinear directions, taking the generic form,
\be
S_\text{multi-jet} =
 \bra{0}  \bar{T} [Y_\bn^\dag{\bf \hat{Y}}^\dag  Y_n(0)]\, \hat{\cM}\,  T [Y_n^\dag {\bf \hat{Y}}  Y_\bn(0)] \ket{0},
\label{eq:Smulti-jet}
\ee
where ${\bf \hat{Y}}$ is a product of outgoing $q, \bar q$, and/or $g$ Wilson lines and $\hat{\cM}$
is an operator that measures momenta of final state particles (see, \eg \cite{Lee:2006nr,Bauer:2008dt}). The directions $n,\bn$ represent the two directions that can flip among $ee,ep,pp$,
thus changing $Y_{n,\bn}$ as in \eq{YinYout}, while  the product ${\bf \hat{Y}}$ remains the same for each.
We consider amplitudes where the same lines are connected by soft gluons, but where the $n$ and/or $\bn$ lines flip direction.
The differences are trivially zero for tree diagrams and for any loop diagrams not involving eikonal propagators from $Y_\bn$ and $Y_n^\dag$. 
The diagrams whose equality for the three processes is nontrivial are loop diagrams involving eikonal propagators
 on one or two of $Y_\bn$ and $Y_n^\dag$, which we call \emph{relevant diagrams}.

In the $\cO(\as)$ multi-jet soft function, the relevant diagrams are essentially the same as ($1\cV$) in \fig{1loop} except that the $n$ and $\bar n$ Wilson lines are replaced by any of the Wilson lines in \eq{Smulti-jet}. They are purely virtual and zero in DR.

Similarly, the relevant diagrams contributing to the $\cO(\as^2)$ multi-jet soft function include all amplitudes in \sec{equality} with $n$ and $\bar n$  lines replaced by $n_1$ and $n_2$ lines, which can be any two lines in \eq{Smulti-jet}. 
(Any diagrams with gluons attached to three or four Wilson lines, as in \fig{34lines}, must contain a purely virtual loop at this order, and, hence, are zero in DR.)  Our arguments in \sec{equality} for most diagrams go through automatically for $n,\bn$ replaced by $n_1,n_2$. The only diagrams for which this generalization is potentially nontrivial are those with the topology of diagrams ($2\cT$) in \fig{1loop}. 
Amplitudes with the topology of ($2\cT$a) with gluons attached to a single Wilson line are still scaleless and zero in DR.
The amplitude with the topology of (2$\cT$b) with gluons attached to Wilson lines $n_{1,2}$ is given by
\begin{align}
\label{eq:Camplitude}
&\cA_{2\cT b}(k;\delta_1,\delta_2) = -\frac{g^3\mu^{3\e}  }{(2\pi)^D}\,  f^{ABC} \,{\bf T}^A_1 {\bf T}^B_2 \ve^C(k) \\
 &\quad\times\int \!\! \frac{d^D q}{q^2 \plus i\e} 
 \frac{1}{(k \minus q)^2 +i\e} \frac{1}{n_1\mcdot (k-q)+i\e\delta_1}\frac{1}{n_2\mcdot q+i\e\delta_2} \nn \\
&\qquad \times 
\biggl[n_1\mcdot (2k- q)\,  n_2 - n_2\mcdot (k+q)\,  n_1 +n_1\mcdot n_2 \,({k}_\perp - 2 {q}_\perp)
      \biggr] \,,\nn
\end{align}
where ${\bf T}^A_i$ is a color charge operator for the $i$th  parton \cite{Catani:1996vz}, which turns into a color matrix $T^A$, $-T^A$, $\cT^A$ for outgoing $q$/incoming $\bar q$,\, outgoing $\bar q$/incoming $q$, outgoing/incoming $g$, respectively.
Unlike \eq{3gluonCamplitude}, we have kept the color factors in \eq{Camplitude} as (potentially) a matrix in color space. 
In the eikonal propagators, the signs $\delta_{i}=\pm$  for outgoing/incoming lines, with the possible combinations 
$(\delta_1,\delta_2)\in \{(+,\, +),\, (+,\, -),\, (-,\, -)\}$, the same as in \eq{3gluonCamplitude}. (A change of variables turns $(-,+)$ back into $(+,-)$.)
It is straightforward to show that the integrals for $(+,+)$ and $(+,-)$ are equal, just as in \eq{3gluonresult}, either by explicit computation or showing that the difference is a scaleless integral in DR, that is,
\be\label{eq:cA124} \!\!
\cA_{2\cT b}(k;+,+)=\cA_{2\cT b}(k;+,-)
\ee
In the $ee$ soft function  $n_{1,2}$ lines are always outgoing and $(\delta_1,\delta_2)_{ee}=(+, +)$.
On the other hand, for $ep$ one of $n_{1,2}$ can be incoming or both can be outgoing, hence
$(\delta_1,\delta_2)_{ep}=(+, \pm)$.
For $pp$, $n_{1,2}$ can be any combination of incoming and outgoing.
Therefore, the difference between amplitudes for
$ee$ and $ep$ is always zero by \eq{cA124}, which immediately implies equality of $ee$ and $ep$ multi-jet soft functions up to $\cO(\as^2)$. 
\be\label{Seeep}
S_{\text{multi-jet}}^{(2)\, ee} = S_{\text{multi-jet}}^{(2)\, ep}
\ee
The non-zero difference between amplitudes relevant for $ep$ vs. $pp$ is similar to \eq{3gluonresult}. We find
\begin{align}
\label{eq:cA23} 
&\cA_{2 \cT b}(k;+,-)-\cA_{2 \cT b}(k;-,-)  =
-\, g^3\mu^{3\e}\,f^{ABC}  \,{\bf T}_1^A {\bf T}_2^B \,
\nn \\
&\quad \times
\frac{[n_1\mcdot k\, n_2 - n_2\mcdot k\,  n_1]\cdot \ve^C(k)}{n_1\mcdot n_2 \,(4\pi)^{1-\e} }
\frac{\Gamma(-\e)^2\Gamma(1+\e)}{\Gamma(-2\e)}
\frac{1}{ \vect{k}_\perp^{2+2\e}} \,.
\end{align}
 This implies the difference between $ep$ and $pp$ amplitudes is nonzero.
To obtain the equality \eq{epppequal} for two legs, we used that the color factor of the amplitude product $\cA_{1R}^\dag \cA_{2\cT b}$ is purely imaginary, and the differing $i\pi$ terms in the integral in \eq{3gluonresult} cancel in the sum over complex conjugates, which proves the equality between $ep$ and $pp$. For the multi-jet result \eq{cA23}, we cannot yet draw the same conclusion in general.
It is possible that the color factor in \eq{cA23} simplifies to be purely imaginary after contracting with the hard function.
 
We can, however, go further for a 3-leg soft function, i.e. with $q\,\bar{q}\,g$ Wilson lines, \eg for $e^+e^-\to\text{3 jets}$, DIS 2-jettiness, or $pp$ 1-jettiness. Then $H$ and $S$ are numbers in a one-dimensional color space, since the only color structure in the hard coefficient is $(C_H)^a_{\alpha \beta}=T^a_{\alpha \beta}$,
where $\alpha\,, \beta$, and $a$ are color indices of the three partons $q^\alpha\,\bar{q}^\beta\,g^a$.
Then, the color factor of the product $\cA_{1R}^{n_3\dag} \cA_{2\cT b}$ multiplied by $C_H$ reduces to 
\be 
f^{ABC} {\bf T}^C_3  {\bf T}^A_1  {\bf T}^B_2 \, C_H =
\begin{cases}
0 \, C_H \quad\quad  \mbox{for } (3,1,2)= ( q, \bar q, g)
\\
 i \left(\tfrac{C_A}{2}\right)^2 C_H \   \mbox{for } ( g,q,g),  (q,g,q)\text{ or }(q\to\bar q)  
\\
i\,\tfrac{C_A}{2}\left(C_F-\tfrac{C_A}{2}\right)  C_H \quad  \mbox{for } (q,q,\bar q), (\bar q,q, \bar q)
\,,
\end{cases}
\label{eq:3jetcolor}
\ee
where  ${\bf T}^C_3$ is the color operator from $\cA_{1R}^{n_3\dag}$. (In \eq{3jetcolor} we assumed the $q,\bar q$ represent outgoing $q,\bar q$.)
These choices and their permutations are all the possible assignments of the lines (3,1,2) to the $q,\bar q,g$ Wilson lines.
Note that cyclic permutations preserve the sign, flipping indices switches it.
(We exclude cases where (3,1,2) are all attached to the same line, which give rise to scaleless diagrams in DR.) Thus $q\bar q g$ soft functions for the same final-state measurement are equal for $e^+e^-$, DIS and DY.

For the case of a soft function with three $ggg$ Wilson lines,  
the color structure of the hard coefficient is $(C_H^{ggg})^a_{bc}=i f^{abc}$.
The color factor corresponding to \eq{3jetcolor} is now $f^{ABC} {\bf T}^C_3  {\bf T}^A_1  {\bf T}^B_2 \, C^{ggg}_H$, which is zero when $(3,1,2)$ are all attached to three different legs and is $\pm i(\tfrac{C_A}{2})^2\,C^{ggg}_H$ when two of $(3,1,2)$ are attached to the same leg, again purely imaginary, the sign depending on the exact placement of the attachments. 
We again exclude the cases when all are attached to the same leg since they give rise to scaleless diagrams in DR.
Therefore, the arguments above still apply, and $ggg$ soft functions are also the same under switching the direction of any Wilson line.

One can perform similar exercises for other multi-jet soft functions once the color structure of the associated hard coefficients is also known.
If their color factors reduce to imaginary numbers, the equality of $ep$ and $pp$ soft functions in these cases is also proved. We leave this explicit check for more than three legs as an open exercise. Even if the color factors turn out not to be imaginary, it would be straightforward to calculate differences between $ep$ and $pp$ soft functions by using \eq{cA23}.

\section{Conclusions}
\label{sec:conclusion}

We have demonstrated that hemisphere soft functions appearing in event shape distributions in $e^+e^-$ collisions, DIS, and Drell-Yan processes are equal in perturbation theory up to $\cO(\as^2)$, which also can be used for non-back-to-back hemisphere event shapes \cite{Kang:2013nha,Stewart:2009yx}. The proof relied on the independence of the final soft functions on the signs of the $\pm i\e$ pole prescriptions in eikonal propagators, which change sign under switching Wilson lines between incoming and outgoing directions. Most amplitudes contributing to the soft functions at this order are transparently independent of these pole prescriptions, with the exception of the one-loop 3-gluon vertex amplitude ($2\cT$b) in \fig{1loop}, or soft gluon current, computed in \cite{Catani:2000pi}, which we reproduced here. For $ee,ep,pp$ soft amplitudes, the real parts are equal, while the imaginary part has opposite sign for $pp$. In the sum over all squared amplitudes including complex conjugates, however, the imaginary parts cancel out, leaving the final soft functions invariant.

While the result for the one-loop soft gluon current was already known \cite{Catani:2000pi}, that it implies the equivalence of $\cO(\as^2)$ soft functions for $ee,ep,pp$ event shape distributions has not been explicitly noticed or exploited before. This observation now allows N$^3$LL resummation for 1-jettiness in DIS  \cite{SCET2014,QCD2014,HERA2014} and 0-jettiness in $pp$. Our proof relied only on the properties of the relevant soft amplitudes, not on details of the measurement function in \eq{measurement}, so the conclusion that soft functions in $ee,ep,pp$ for the same measurement function are equal to $\cO(\as^2)$ is quite general. We also showed that $\cO(\as^2)$ soft functions with three legs also obey the same equivalence properties under switching lines between incoming and outgoing. Some other soft functions computed to $\cO(\as^2)$ for which this equivalence should hold include: transverse-momentum dependent distributions in DY \cite{Li:2011zp}, jet mass with a jet veto \cite{Kelley:2011aa}, DY threshold resummation \cite{Belitsky:1998tc,Becher:2007ty} and Higgs threshold resummation \cite{Ahrens:2008nc}, jet broadening \cite{Becher:2012qc}, soft functions with three Wilson lines such as in $pp\to H+\text{jet}$ \cite{Becher:2012za}, and more.

\emph{Note added:} As this paper was being completed, Ref.~\cite{Boughezal:2015eha} appeared, presenting a framework for computing $N$-jettiness soft functions to $\cO(\as^2)$ numerically. It included the analytic $\cO(\as^2)$ DIS 1-jettiness soft function, obtained from the $e^+e^-$ soft function computed in \cite{Monni:2011gb}, in agreement with our proof of their equivalence, but without the proof made explicit. Our proof also implies their equivalence with the $\cO(\as^2)$ $pp$ 0-jettiness soft function. Our results, in particular on multi-jet soft functions in \sec{multi-jet}, also imply that the $pp$ 1-jettiness soft function in \cite{Boughezal:2015eha} would remain the same at $\cO(\as^2)$ under changes of the directions of any of the Wilson lines from incoming to outgoing.

\section*{Acknowledgements}
We thank J. Collins, A. Hornig and I. Stewart for helpful discussions and feedback. OZL thanks Los Alamos National Laboratory for hospitality during the course of this work. The work of DK and CL is supported by DOE Contract DE-AC52-06NA25396 and by the LDRD program at LANL. The work of OZL is supported by DOE Grant No. DE-FG02-04ER41338.

\appendix

\setcounter{figure}{0}

\section{Known perturbative results to $\cO(\as^2)$}
\label{appx:pert}

Here we give known results for the pieces of the $\cO(\as^2)$ hemisphere soft function in \eq{S2terms}.
The first set of terms $R_c$ can be deduced from the known soft anomalous dimension \cite{Hoang:2008fs},
\begin{align}
\label{eq:Rc2}
&R_c^{(2)}(\ell_1,\ell_2,\mu) = 2 C_F^2 (L_1^4 + L_2^4) + 4 C_F^2 L_1^2 L_2^2 \\
&+ \Bigl( \frac{22}{9}C_F C_A - \frac{8}{9} C_F T_R n_f\Bigr) (L_1^3 + L_2^3) + \Bigl[ - \frac{5\pi^2}{3} C_F^2 \nn \\
&+ C_F C_A \Bigl( \frac{\pi^2}{3} - \frac{67}{9}\Bigr) + \frac{20}{9} C_F T_R n_f\Bigr] (L_1^2 + L_2^2) \nn\\
&+ \Bigl[ 16\zeta_3 C_F^2 + C_FC_A \Bigl( \frac{202}{27} - \frac{11\pi^2}{18} - 7\zeta_3\Bigr)\nn \\
&\quad - C_F T_R n_f \Bigl( \frac{56}{27} - \frac{2\pi^2}{9}\Bigr)\Bigr] (L_1 + L_2) 
- C_F^2 \frac{7\pi^4}{45}\nn \\
& + C_F C_A \Bigl( \frac{88\zeta_3}{9} + \frac{67\pi^2}{27} - \frac{\pi^4}{9}\Bigr) - C_F T_R n_f \Bigl(\frac{32\zeta_3}{9} + \frac{20\pi^2}{27}\Bigr)
\,, \nn
\end{align}
where $L_{1,2} = \ln (\ell_{1,2}/\mu)$. 

The result from \cite{Hornig:2011iu} for the non-constant non-global terms $S_{\text{NG}}$ terms can be expressed
\begin{align}
\label{eq:SNG2}
&S_{\text{NG}}^{(2)} (\ell_1,\ell_2)= - \frac{\pi^2}{3} C_F C_A \ln^2\frac{\ell_1}{\ell_2} \\
&+ \Bigl( C_F C_A \frac{11\pi^2 \minus 3 \minus 18\zeta_3}{9} + C_F T_R n_f \frac{6-4\pi^2}{9} \Bigr) \ln\frac{\ell_1/\ell_2 \plus \ell_2/\ell_1}{2} \nn \\
&+ C_F C_A \Bigl[ f_N\Bigl(\frac{\ell_1}{\ell_2}\Bigr) + f_N\Bigl(\frac{\ell_2}{\ell_1}\Bigr) - 2f_N(1)\Bigr] \nn \\
&+ C_F T_R n_f \Bigl[ f_Q\Bigl(\frac{\ell_1}{\ell_2}\Bigr) + f_Q\Bigl(\frac{\ell_2}{\ell_1}\Bigr) - 2f_Q(1)\Bigr]\,, \nn
\end{align}
where the functions $f_{N,Q}$ are given by
\begin{align}
\label{eq:fQN}
&f_Q(a) = \Bigl[ \frac{2\pi^2}{9} - \frac{2}{3(a+1)}\Bigr] \ln a - \frac{4}{3}\ln a \Li_2(-a) + 4\Li_3(-a) \nn \\
&\quad+ \frac{2\pi^2-3}{9} \ln\Bigl(a+\frac{1}{a}\Bigr)\,, \nn \\
&f_N(a) = -4\Li_4\Bigl(\frac{1}{a+1}\Bigr) - 11\Li_3(-a) + 2\Li_3\Bigl(\frac{1}{a+1}\Bigr) \ln\frac{a}{(a+1)^2} \nn \\
&\ + \Li_2\Bigl(\frac{1}{a+1}\Bigr) \Bigl[ \pi^2 - \ln^2(a+1) - \frac{1}{2}\ln a \ln\frac{a}{(a+1)^2} + \frac{11}{3}\ln a\Bigr] \nn \\
&\ + \Bigl[ \frac{11}{12}\ln\frac{a}{(a+1)^2} - \frac{1}{4} \ln\frac{a+1}{a} \ln(a+1) + \frac{\pi^2}{24}\Bigr] \ln^2 a \nn \\
&\ - \frac{1}{6}\frac{a-1}{a+1} \ln a + \frac{5\pi^2}{12}\ln\frac{a+1}{a} \ln (a+1) - \frac{11\pi^4}{180} \nn \\
&\ - \frac{11\pi^2 -3 -18\zeta_3}{18} \ln\Bigl( a + \frac{1}{a}\Bigr)\,.
\end{align}
These functions are bounded and vanish as $a\to0,\infty$. Their values at $a=1$ are
\be
\begin{split}
2f_Q(1) &= -6\zeta_3 + \frac{2}{9}(2\pi^2 -3)\ln 2 \\
2f_N(1) &= -8\Li_4\frac{1}{2} + \zeta_3\Bigl(\frac{33}{2}-5\ln2\Bigr) + \frac{\ln 2 - \ln^4 2}{3} \\
&\quad + \frac{2\pi^4}{45} + \frac{\pi^2}{3} \Bigl(\ln^2 2 - \frac{11}{3}\ln2\Bigr)\,,
\end{split}
\ee
and are subtracted out of the last two lines of \eq{SNG2} so that $S_{\text{NG}}$ vanishes at $\ell_1=\ell_2$.

The constant term $c_S^{(2)}$ was computed in \cite{Kelley:2011ng,Monni:2011gb}, with the result
\begin{align}
\label{eq:cS2}
c_S^{(2)} &= C_F^2 \frac{\pi^4}{8} + C_F C_A \biggl[ - \frac{508}{81} - \frac{871}{216}\pi^2 + \frac{4\pi^4}{9} + \frac{22}{9} \zeta_3 \nn \\
&\quad  - 7\zeta_3 \ln2 + \frac{\pi^2}{3} \ln^2 2 - \frac{1}{3}\ln^4 2 - 8\Li_4\Bigl(\frac{1}{2}\Bigr) \biggr] \nn \\
&\quad + C_F T_R n_f \biggl( -\frac{34}{81} + \frac{77}{54}\pi^2 - \frac{8}{9}\zeta_3 \biggr) \,.
\end{align}
Thus the final result for the $\cO(\as^2)$ hemisphere soft function in $e^+e^-$ is given by \eq{S2terms} with the three individual pieces given by \eqss{Rc2}{SNG2}{cS2}.

The position-space soft function is defined by the Fourier transform of the momentum space \eq{softamplitudes}, and takes a form analogous to \eq{S2terms}.
All the non-constant terms at $\cO(\as^2)$ were computed in \cite{Hornig:2011iu}. The constants at $\cO(\as^2)$ can be obtained analytically from the momentum-space results of \cite{Kelley:2011ng}, giving
\begin{align}
\tilde c_S^{(2)} &= C_F^2 \frac{\pi^4}{8} + C_F C_A \Bigl(-\frac{535}{81} - \frac{871}{216}\pi^2 + \frac{7}{30}\pi^4 + \frac{143}{18}\zeta_3\Bigr) \nn \\ 
&\quad + C_F T_R n_f \Bigl( \frac{20}{81} + \frac{77}{54}\pi^2 - \frac{26}{9}\zeta_3\Bigr)\,.
\end{align}

In this section we have reviewed the previously known results for the $e^+e^-$ hemisphere soft function at  $\cO(\as^2)$, which we have shown in this paper is also equal to those for DIS and $pp$.

\section{Three-gluon vertex diagram for $ep,pp$}
\label{appx:threegluon}

In this Appendix we provide an explicit computation of the amplitude ($2\cT$b) in \fig{1loop}, the result of which is given by \eqs{3gluonCnonzero}{3gluonresult}, for the $ee,ep,pp$ soft functions. This will reproduce the result for the soft gluon current at one loop given in \cite{Catani:2000pi}, but we will find it instructive to provide our own derivation, showing in particular how the $i\pi$ term in \eq{3gluonresult} arises. 

\begin{figure}
\begin{center}
\includegraphics[width=0.37\textwidth]{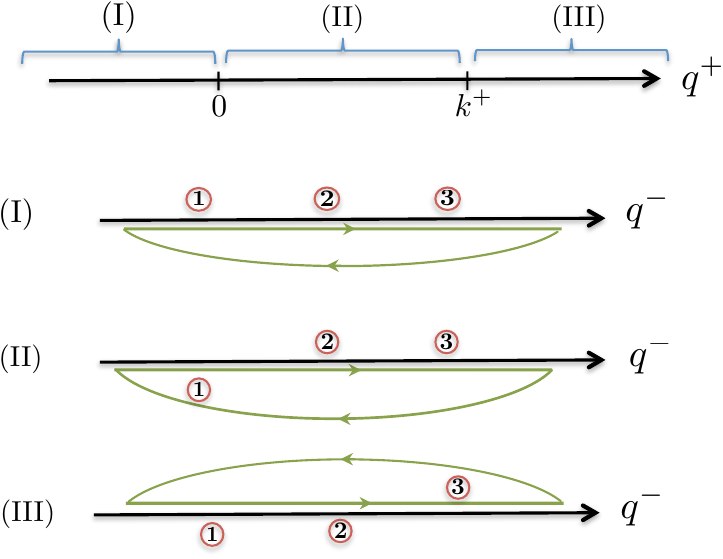}
\end{center}
\vspace{-1em}
\caption{Positions of the three complex $q^-$ poles of $\cI_\cT^{ep,pp}$ in \eq{qminuspoles}, as a function of $q^+$. In region I where $q^+<0$, the $q^-$ contour can be closed below the real axis, giving zero for the integral \eq{ICee}, while for $q^+>0$ in regions II and III the contour is closed below or above the real axis as shown, yielding the result in \eq{IepIppcontour}.}
\label{fig:pp_ep_q- pole}
\end{figure}

The $\cI_\cT$ integrands in \eq{ICee} for $ee,ep$ have the same pole structure in $q^+$, while for $ep$ and $pp$ they have the same pole structure in $q^-$. Namely, $\cI_\cT^{ee,ep}$ have poles in $q^+$ at:
\be
\label{eq:qpluspoles}
q^+ = k^+ + i\e\,, \frac{\vect{q}_\perp^2 -i\e}{q^-} \,, \frac{\vect{q}_\perp^2 - 2\vect{q}_\perp\mcdot\vect{k}_\perp + q^- k^+ - i\e}{q^- - k^-}\,,
\ee
whiile $\cI_\cT^{ep,pp}$ have poles in $q^-$ at:
\be
\label{eq:qminuspoles}
q^- =  i\e\,, \frac{\vect{q}_\perp^2 -i\e}{q^+} \,, \frac{\vect{q}_\perp^2 - 2\vect{q}_\perp\mcdot\vect{k}_\perp + q^+ k^- - i\e}{q^+ - k^+}\,,
\ee
labeled 1, 2, and 3, respectively, in \fig{pp_ep_q- pole}, which illustrates the position of the $q^-$ poles in \eq{qminuspoles} in the upper- or lower-half complex $q^-$ plane as a function of $q^+$.
Here we will perform the $\cI_\cT^{ep,pp}$ integrals explicitly by contour integration in $q^-$. The computation of $\cI_\cT^{ee,ep}$ is similar, but we will not give the details here. Both calculations yield the result in \eq{3gluonresult}.

Performing this $q^-$ contour integration for $\cI_\cT^{ep,pp}$ in \eq{ICee}, for $q^+<0$ (region I in \fig{pp_ep_q- pole}), we can close the contour in the lower half plane and obtain zero. The nonzero contributions come from the other two regions, II and III in \fig{pp_ep_q- pole},
\be
\label{eq:IepIppcontour}
\cI_\cT^{ep,pp}(k) = -\frac{i}{4\pi} \int\frac{d^dq_\perp}{(2\pi)^d} \frac{1}{\vect{q}_\perp^2} \int_0^\infty dq^+ \frac{ F(q^+,\vect{q}_\perp,k)}{q^+ - k^+ \mp i\e}\,,
\ee
where $d = 2-2\e$,  the upper (lower) signs in $\mp i\e$ in the $q^+$ eikonal propagator are for $ep$ ($pp$), and $F$ is given by
\be
\label{eq:Feppp}
F(q^+,\vect{q}_\perp,k) \equiv \frac{q^+}{k^+} \frac{\theta(k^+ - q^+)}{\bigl( \vect{q}_\perp - \frac{q^+}{k^+}\vect{k}_\perp\bigr)^2} \\
+ \frac{\theta(q^+ - k^+) }{\bigl(\vect{q}_\perp - \vect{k}_\perp)^2 + \vect{k}_\perp^2 (\frac{q^+}{k^+} - 1)}\,.
\ee
We used the on-shell condition $k^- = \vect{k}_\perp^2/k^+$ to eliminate $k^-$ from this expression. 
The $i\e$'s in the $\vect{q}_\perp$-dependent propagators in \eqs{IepIppcontour}{Feppp} can be dropped since the denominators are $\geq 0$, and the integral over $\vect{q}_\perp$ does not cross over any singularities. The $q^+$ integral in \eq{IepIppcontour}, however, goes over the singularity at $q^+ = k^+$, and we use the prescription
\be
\label{eq:PVprescription}
\frac{1}{q^+ - k^+ \mp i\e} = \PV\frac{1}{q^+-k^+} \pm i\pi \delta(q^+ - k^+)
\ee
to perform the integral. The function $F$ is finite and continuous at $q^+ = k^+$:
\be
F(k^+,\vect{q}_\perp,k) = \frac{1}{(\vect{q}_\perp - \vect{k}_\perp)^2}\,.
\ee
The result of using this prescription in \eq{IepIppcontour} can be expressed
\be
\label{eq:IepIppAB}
\cI_{ep,pp}(k) = -\frac{i}{4\pi} A \pm \frac{1}{4} B\,,
\ee
where
\begin{subequations}
\begin{align}
A &\equiv  \int \frac{d^d q_\perp}{(2\pi)^d} \frac{1}{\vect{q}_\perp^2} \int_0^\infty dq^+ F(q^+,\vect{q}_\perp,k)  \PV \frac{1}{q^+-k^+}  \\
B &\equiv \int \frac{d^d q_\perp}{(2\pi)^d} \frac{1}{\vect{q}_\perp^2 (\vect{q}_\perp - \vect{k}_\perp)^2}\,.
\end{align}
\end{subequations}
$B$ is easily evaluated. Combining denominators using a Feynman parameter and then completing the integrations, we obtain the result
\be
B = \frac{1}{(4\pi)^{1-\e}} \frac{\Gamma(1+\e) B(-\e,-\e)}{(\vect{k}_\perp^2)^{1+\e}}\,,
\ee
where $B(a,b)$ is the beta function. 

To evaluate $A$ we must regulate the singularity at $q^+ = k^+$ consistently with the principal value prescription. This can be done with symmetric cutoffs around $q^+ = k^+$, or, conveniently, we can insert a factor (similar to, but not directly associated with, the rapidity regulator in \SCETb \cite{Chiu:2011qc,Chiu:2012ir}):
\be
\label{eq:Aeta}\!\!
A = \lim_{\eta\to 0} \int\frac{d^dq_\perp}{(2\pi)^d} \frac{1}{\vect{q}_\perp^2} \int_0^\infty\!\! dq^+ \biggl(\frac{\nu}{\abs{q^+-k^+}}\biggr)^\eta \frac{F(q^+,\vect{q}_\perp,k)}{q^+-k^+}\,.
\ee
Using the changes of variables $q^+ \to q'= \abs{k^+ - q^+}$ and then $q' = k^+ u$,
and combining the $\vect{q}_\perp$ denominators in \eqs{Aeta}{Feppp} using a Feynman parameter, we obtain as the result of performing the $\vect{q}_\perp$ integral,
\be
\begin{split}
A &= \frac{1}{(4\pi)^{1-\e}} \frac{\Gamma(1+\e)}{ (\vect{k}_\perp^2)^{1+\e}} \biggl(\frac{\nu}{k^+}\biggr)^\eta \int_0^1 \frac{dx}{x^{1+\e}} \\
&\qquad\times \biggl\{ -\int_0^1 \frac{du}{u^{1+\eta}}\, \frac{1}{(1-x)^{1+\e} (1-u)^{1+2\e}} \\
&\qquad\qquad + \int_0^\infty \frac{du}{u^{1+\eta}}\, \frac{1}{x^{1+\e}  (1-x+u)^{1+\e}}  \biggr\}\,.
\end{split}
\ee
The two $u$ integrals have $1/\eta$ poles, but they cancel, and we can take the $\eta\to 0$ limit to obtain
\be
A = \frac{1}{(4\pi)^{1-\e}} \frac{\Gamma(1+\e)}{ (\vect{k}_\perp^2)^{1+\e}}B(-\e,-\e) \frac{\pi}{\tan(\pi\e)}\,,
\ee
Thus the sum of $A,B$ terms in the integral \eq{IepIppAB} yields
\be
\label{eq:IepIppresult}
\begin{split}
\cI_{ep,pp} &= -\frac{i}{16\pi^2} (4\pi)^\e\frac{\Gamma(1+\e)}{ (\vect{k}_\perp^2)^{1+\e}}B(-\e,-\e)  \frac{\pi e^{\pm i\pi\e}}{\sin(\pi\e)}\,,
\end{split}
\ee
consistent with the result for the one-loop soft gluon current in \cite{Catani:2000pi}.
Plugging this integral back into the amplitude \eq{3gluonCnonzero}, multiplying by the sum of conjugates of the 1-gluon tree-level amplitudes from \eq{eikonalamplitudes}, and summing over final-state polarizations and integrating over the final-state gluon momentum $k$ in \eq{softamplitudes}, we obtain for this contribution to the soft function,
\begin{align}
&S_2^{ep,pp} = \frac{1}{N_C}\Tr \int\frac{d^D k}{(2\pi)^D} 2\pi\delta(k^2) \theta(k^0) \cM_{\ell_1\ell_2}(k) \\
&\qquad\qquad\qquad\qquad\qquad \times  \cA_{ep,pp}^\cT(k) [\cA_{1n}^\dag(k) + \cA_{1\bn}^\dag(k)] \nn \\
&\quad=\frac{\as^2\, C_A C_F }{16\pi^2}\mu^{4\e} \left[ \frac{\delta(\ell_2)}{\ell_1^{1+4\e} } +\frac{\delta(\ell_1)}{\ell_2^{1+4\e}}  \right] \nn\\
&\quad \times \frac{1}{\e}\biggl\{ - \frac{2}{\e^2} + \pi^2 + \frac{16\zeta_3}{3} \e - \frac{\pi^4}{60}\e^2 
\pm i\pi \biggl(-\frac{2}{\e} + \frac{\pi^2}{3}\e + \frac{16\zeta_3}{3}\e^2\biggr)\biggr\}  \,.\nn
\end{align}
Upon adding the complex conjugate diagrams, $S_2^{ep,pp} + S_2^{ep,pp*}$, the imaginary parts cancel and the real parts combine to reproduce the result for these diagrams in $S_2^{ee}$ given in \cite{Kelley:2011ng}.

Some similar features of the loop integrals in diagrams of similar topology as the one computed here were observed in the computation of the gluon beam function in \cite{Gaunt:2014cfa}.

  \bibliographystyle{apsrev} 
  \bibliography{soft}


%
%
%
\end{document}